\documentclass[amsmath,amssymb,twocolumn,nofootinbib]{revtex4-1}
\usepackage{graphicx}
\usepackage{epsfig}
\begin{document}
\title{Induced gravity, and minimally and conformally coupled scalar fields  in Bianchi-I cosmological models}
\author{Alexander~Yu.~Kamenshchik}
\email{kamenshchik@bo.infn.it}
\affiliation{Dipartimento di Fisica e Astronomia, Universit\`a di Bologna\\ and INFN,  Via Irnerio 46, 40126 Bologna,
Italy,\\
L.D. Landau Institute for Theoretical Physics of the Russian
Academy of Sciences,\\
 Kosygin str. 2, 119334 Moscow, Russia}
\author{Ekaterina~O.~Pozdeeva}
\email{pozdeeva@www-hep.sinp.msu.ru}
\affiliation{Skobeltsyn Institute of Nuclear Physics, Lomonosov Moscow State University,\\
 Leninskie Gory 1, 119991, Moscow, Russia}
 \author{Alexei A. Starobinsky}
 \email{alstar@landau.ac.ru}
\affiliation{L.D. Landau Institute for Theoretical Physics of the Russian Academy of Sciences,\\
 Kosygin str. 2, 119334 Moscow, Russia\\
 Kazan Federal University, Kazan 420008, Republic of
Tatarstan, Russia}
\author{Alessandro~Tronconi}
\email{tronconi@bo.infn.it}
\affiliation{Dipartimento di Fisica e Astronomia, Universit\`a di Bologna\\ and INFN, Via Irnerio 46, 40126 Bologna,
Italy}
\author{Giovanni~Venturi}
\email{giovanni.venturi@bo.infn.it}
\affiliation{Dipartimento di Fisica e Astronomia, Universit\`a di Bologna\\ and INFN,  Via Irnerio 46, 40126 Bologna,
Italy}
\author{Sergey~Yu.~Vernov}
\email{svernov@theory.sinp.msu.ru}
\affiliation{Skobeltsyn Institute of Nuclear Physics, Lomonosov Moscow State University,\\
 Leninskie Gory 1, 119991, Moscow, Russia}

\begin{abstract}
We study the cosmological evolution and singularity crossing in the Bianchi-I universe filled with a conformally coupled scalar field and compare them
with those of the Bianchi-I universe filled with a minimally coupled scalar field.
 We also write down  the solution for the Bianchi-I Universe in the induced gravity cosmology.  
\end{abstract}
\maketitle

\section{Introduction}

The cosmological singularity problem has attracted  the  attention of researchers working in general relativity and its modifications for a long time. 
Already in the seminal paper by Robertson \cite{Robertson}, where the early development of the Friedmann-type cosmologies was reviewed and 
generalized, the question of the initial singularity was discussed.  The connection between the presence and sign of the spatial curvature, the value 
of the cosmological constant and the character of the dependence of the pressure on the scale factor of the universe and the appearance of a singularity
were studied there in detail. It is interesting that Robertson also considered the scenario of the cyclic evolution of the universe according to some trigonometrical 
law, when it exits of the singularity expands until some maximal value of its radius then contracts and the the process repeats itself indefinitely.  
It appears that the fact the universe passes through this singularity did not disturb him too much.  

Later, the question of the generality of the initial cosmological singularity and its appearance not only in the homogeneous and isotropic Friedmann models
was intensively discussed \cite{Khal-Lif,Pen-Hawk}. The general theorems about the existence of such a singularity were proven \cite{Pen-Hawk} and 
the oscillatory approach to the cosmological singularity \cite{BKL}, known also as Mixmaster Universe \cite{Misner}, was discovered as a result of these discussions.  

Another type of cosmological singularities, which can arise in the future for some finite values of the scale factor of the universe 
and can be rather soft was described in paper \cite{Barrow-soft}. The interest in such  singularities essentially increased during last years (see e.g. \cite{soft,soft1,
soft2,we-tach,branes,my-rep}). At this point the condition for the crossing of such singularities become important. Such conditions were analyzed  in detail in paper 
\cite{Lazkoz}, where the opportunity of continuation of the geodesics through them was studied. Some curious effects arising at the crossing of the soft future singularities were described, for example, in paper \cite{we-cross}.

In contrast with the crossing of the soft singularities the idea of the possible  crossing of the Big Bang -- Big Crunch singularity appears rather counter-intuitive. For many years the desire of looking for models free of such singularities  dominated. However, in some cosmological models the idea of the possible transition from big crunch to big bang was 
studied. First of all, we would like to mention the string or pre-big bang scenario \cite{Gasp-Ven,Lidsey,Gasp-Ven1}. As is well known a definite prediction of string theory is the existence of a scalar field referred to as the dilaton, which couples directly to matter \cite{Scherk}. Its vacuum expectation value determines the strengths of both the gauge and gravitational couplings. In the pre-big bang cosmological model, the accelerated expansion of the universe is driven by the kinetic energy of the dilaton field.The presence of this dilaton field is essential also from the point of view of discussion of the cosmological singularity problem. Indeed, it was noticed in the framework of the string pre-big bang cosmology that the transition from the string frame, where the dilaton is non-trivially coupled to gravity to the Einstein frame can change drastically the observable evolution of the universe and what looks like an expansion in one frame can look like a contraction in another frame.  We would like also to note that in the framework of the string pre-big bang cosmology  non only isotropic Friedmann models, but also anisotropic Bianchi-I models \cite{Copeland,Copeland1} were studied.
In these works the universe was filled not only by the dilaton, but also by an antisymmetric tensor field, which influenced its dynamics.

Another approach to the problem of the singularity also inspired by the superstring theories was developed in papers \cite{Khoury, Khoury1, Khoury2}. 
In paper \cite{Khoury}  a cosmological scenario was proposed, in which the hot big bang universe was produced by the collision of a brane in the bulk space with a bounding orbifold plane, beginning from an otherwise cold, vacuous, static universe. A particularly interesting aspect of this cosmological scenario is that some characteristics of the universe suggest slow expansion, other superluminal expansion, other - contraction. This is possible due to the fact that for different equations the role of the scale factor is played by different combinations of moduli fields. Some combinations increase with time, mimicking an expanding universe, and others decrease, mimicking a contracting universe.  In paper \cite{Khoury1} the authors treat the singularity as transition between a contracting big crunch phase and an expanding big bang phase.  A crucial role in their analysis is played by a massless scalar field - a modulus. The theory is reformulated in such a way as to define variables, which are finite as the scale factor shrinks to zero. That suggests a natural way to match the solutions before and after the singularity. A general scenario of cyclic universe passing through the singularity is presented 
in paper \cite{Khoury2}. The general features of the approach \cite{Khoury,Khoury1,Khoury2} are the  role of the scalar field and the construction of variables which are finite 
at the singularity crossing. These features are essential also for the approach which we develop in this paper and in the preceding papers \cite{KPTVV-sing,we-Bianchi}
However before presentation of our treatment we would like to mention some other works. 
During the last decade some approaches to the problem of the description of such a crossing  were elaborated \cite{Bars,Bars1,Wetterich,Wetterich1,Prester,Prester1}. Behind  these approaches there are basically two general ideas. Firstly, to cross the singularity one must  give a prescription matching non-singular, finite quantities before and after such a crossing. Secondly, such a description can be achieved by using a convenient choice of  field parametrization.

In our preceding paper \cite{KPTVV-sing}  we   proposed a  version of the description of the crossing of singularities  in  universes filled with scalar fields. This version was based on the transitions between the Jordan and the Einstein frames.  In \cite{KPTVV-sing}  we   only considered  an isotropic cosmological singularity, present in a flat Friedmann universe. We also  essentially used the relations between exact solutions of the cosmological (Friedmann and Klein-Gordon) equations in two different frames, which were studied in detail in our papers \cite{KPTVV2013,KPTVV2015}.  The main idea of
  the paper \cite{KPTVV-sing} was the following: when in the Einstein frame the universe arrives to the Big Bang -- Big Crunch singularity, from the point of view of the evolution of its counterpart in the Jordan frame its geometry is regular, but the effective Planck mass has a zero value. The solution to the equations of motion in the Jordan frame is smooth at this point and on using the relations between the solutions of the cosmological equations in the two frames one can describe the crossing of the cosmological singularity in a uniquely determined way. The contraction is replaced by the expansion (or vice versa)
and the universe enters into the antigravity regime. Analogously, when the geometry is singular in the Jordan frame it is regular in the Einstein frame, and on using this regularity we can describe in a well determined way the crossing of the singularity in the Jordan frame.

We would like also to mention the series of papers \cite{Bron1,Bron2,Bron3,Bron4} where the relation between the exact static solutions in the Jordan frame and in the Einstein frame was studied in detail. In particular, it was noticed that the singularity in one frame can correspond to the regular surface in another frame. 

 It is important to note that the possibility of a change of sign of the effective gravitational  constant and of the construction of a singularity-free isotropic cosmological models including a scalar field conformally coupled to the scalar curvature was studied in papers \cite{Bekenstein, Grib,Frolov,Melnikov1,Melnikov,Linde1979}.
In paper \cite{Starobinsky1981} this possibility was analyzed in detail.

 Namely, in paper \cite{Starobinsky1981} it was pointed out  that in a homogeneous and isotropic universe one can indeed  cross the point where the effective gravitational constant changes sign. 
However, the presence of anisotropies or inhomogeneities changes the situation drastically, because when the value of the effective Planck mass tends to zero, and hence, the effective  gravitational constant tends to infinity, these anisotropies and inhomogeneities grow indefinitely.

In our paper \cite{we-Bianchi} we have investigated this phenomenon, suggesting a simple field reparametrization,
allowing one to describe the Big Bang -- Big Crunch singularity crossing in the Bianchi-I model filled with a minimally coupled scalar field.     In paper \cite{we-Bianchi} we  also  wrote down  asymptotic solutions for the Bianchi-I universe filled with a conformally coupled scalar field, describing its behavior just after the Big Bang and in the far future.  However,  an exact solution for this model on using a particular time parametrization was obtained many years ago \cite{Star-dipl}. In the present paper we study the properties of this old solution \cite{Star-dipl}, which is very interesting from the point of view of the possibility of the description of the singularity crossing. We also compare the cosmological evolutions in the  Bianchi-I universe, filled with a minimally coupled or with a conformally coupled scalar fields and see that such evolutions are qualitatively different. Further, we obtain an exact solution for the Bianchi - I universe in the induced gravity theory. 

The structure of the paper is as follows: in the second section we present some general formulas for  gravity with a conformally coupled scalar field. Section III is devoted to the exact solution for the Bianchi-I model filled with a conformally coupled scalar field. In the fourth section we describe the complete evolution and the crossing of the singularities in this model and we also compare this evolution with that in the Bianchi-I universe, filled with a minimally coupled scalar field.  In the fifth section we obtain
the general exact solution for a Bianchi - I universe in induced gravity theory and the last section contains a brief concluding remark.

\section{Some general formulas for  gravity with a conformally coupled scalar field}

Let us consider an action
\begin{equation}
S = \int\sqrt{-g}\left(U(\sigma)R - \frac12g^{\mu\nu}\sigma_{,\mu}\sigma_{,\nu} + V(\sigma)\right),
\label{action}
\end{equation}
where $\sigma$ is a scalar field and the curvature scalar is
$R = g^{\mu\nu}R_{\mu\nu}$.
The curvature tensor is defined as
\begin{equation}
R^{\alpha}_{\mu\beta\nu} = \frac{\partial \Gamma^{\alpha}_{\mu\nu}}{\partial x^{\beta}} - \frac{\partial \Gamma^{\alpha}_{\mu\beta}}{\partial x^{\nu}} + \Gamma^{\alpha}_{\gamma\beta}\Gamma^{\gamma}_{\mu\nu} - \Gamma^{\alpha}_{\gamma\nu}\Gamma^{\gamma}_{\mu\beta},
\label{curvature}
\end{equation}
while the Ricci tensor is
\begin{equation}
R_{\mu\nu}=R^{\alpha}_{\mu\alpha\nu} = \frac{\partial \Gamma^{\alpha}_{\mu\nu}}{\partial x^{\alpha}} - \frac{\partial \Gamma^{\alpha}_{\mu\alpha}}{\partial x^{\nu}} + \Gamma^{\alpha}_{\gamma\alpha}\Gamma^{\gamma}_{\mu\nu} - \Gamma^{\alpha}_{\gamma\nu}\Gamma^{\gamma}_{\mu\alpha}.
\label{curvature1}
\end{equation}

Using the standard formula expressing the Christoffel symbols $\Gamma^{\alpha}_{\mu\nu}$ via the metric tensor, we get the variation of the Ricci tensor with respect to the metric:
\begin{eqnarray}
&&\delta R_{\alpha\beta} = \frac12g^{\gamma\delta}(\nabla_{\gamma}\nabla_{\beta}(\delta g_{\delta\alpha})+\nabla_{\gamma}\nabla_{\alpha}(\delta g_{\delta\beta})\nonumber \\
&&-\nabla_{\gamma}\nabla_{\delta}(\delta g_{\alpha\beta})-\nabla_{\beta}\nabla_{\alpha}(\delta g_{\gamma\delta})).
\label{variation}
\end{eqnarray}

Using Eq.~(\ref{variation}), one can show that after  integration by parts, the term
\begin{equation*}
U(\sigma)g^{{\mu\nu}}\delta R_{{\mu\nu}}
\end{equation*}
in the variation of the action (\ref{action}) will be transformed into
\begin{equation*}
(g_{\mu\nu}\Box U -\nabla_{\mu}\nabla_{\nu}U)\delta g^{\mu\nu}.
\end{equation*}

By this way, varying the action (\ref{action}) with respect to the contravariant metric components, we obtained
the Einstein equation
\begin{eqnarray}
&&U\left(R_{\mu\nu}-\frac12g_{\mu\nu}R\right)+g_{\mu\nu}\Box U-\nabla_{\mu}\nabla_{\nu} U \nonumber \\
&&=\frac12\sigma_{,\mu}\sigma_{,\nu}-\frac14g_{\mu\nu}\sigma_{,\alpha}\sigma^{,\alpha}+\frac12g_{\mu\nu}V.
\label{Einstein}
\end{eqnarray}

The variation with respect to $\sigma$ gives the Klein-Gordon equation
\begin{equation}
\Box \sigma +V' + U' R =0,
\label{KG}
\end{equation}
where ``prime'' means  differentiation with respect to $\sigma$.

Substituting the expressions
\begin{eqnarray*}
&&\nabla_{\mu}U=U' \sigma_{,\mu},\nonumber \\
&&\nabla_{\nu}\nabla_{\mu} U = U''\sigma_{,\mu}\sigma_{,\nu}+U'\nabla_{\nu}\nabla_{\mu}\sigma,\nonumber \\
&&\Box U = U''\sigma_{,\alpha}\sigma^{,\alpha}+U'\Box \sigma.
\end{eqnarray*}
into Eq.~(\ref{Einstein}), we obtain
\begin{eqnarray}
&&U\left(R_{\mu\nu}-\frac12g_{\mu\nu}R\right)\nonumber\\
&&+g_{\mu\nu}U''\sigma_{,\alpha}\sigma^{,\alpha}+g_{\mu\nu}U'\Box \sigma-U''\sigma_{,\mu}\sigma_{,\nu}-U'\nabla_{\nu}\nabla_{\mu}\sigma \nonumber \\
&&=\frac12\sigma_{,\mu}\sigma_{,\nu}-\frac14g_{\mu\nu}\sigma_{,\alpha}\sigma^{,\alpha}+\frac12g_{\mu\nu}V.
\label{Einstein1}
\end{eqnarray}
Contracting Eq.~(\ref{Einstein1}) with the contravariant metric, we get
\begin{eqnarray}
&&-UR+3U''\sigma_{,\mu}\sigma^{,\mu}+3U'\Box\sigma
-\frac12\sigma_{,\mu}\sigma^{,\mu}+2V.
\label{Einstein2}
\end{eqnarray}
In the case of the conformal coupling
\begin{equation}
U=U_0-\frac{\sigma^2}{12}
\label{conformal}
\end{equation}
Eq.~(\ref{Einstein2}) is reduced to
\begin{equation}
\left(U_0-\frac{\sigma^2}{12}\right)R+\frac12\sigma\Box \sigma+2V=0,
\label{Einstein3}
\end{equation}
while the Klein-Gordon equation (\ref{KG}) becomes
\begin{equation}
\Box\sigma = -V'+\frac{\sigma}{6}R.
\label{KG1}
\end{equation}

Substituting the expression for $\Box\sigma$ from Eq.~(\ref{KG1}) into Eq.~(\ref{Einstein3}), we obtain
\begin{equation}
U_0R + 2V - \frac12\sigma V'=0.
\label{Einstein4}
\end{equation}

Thus, if $V=0$, then $R=0$, and the Klein-Gordon equation is simply
\begin{equation}
\Box \sigma =0.
\label{KG2}
\end{equation}

\section{Bianchi-I cosmological model}

We consider the Bianchi-I cosmological  model with a metric
\begin{eqnarray}
ds^2 &=& N^2(\tau)d\tau^2 - a^2(\tau)\left(e^{2\beta_1(\tau)}(dx^1)^2\right.\nonumber \\
 &+&\left. e^{2\beta_2(\tau)}(dx^2)^2+e^{2\beta_3(\tau)}(dx^3)^2\right),
 \label{Bianchi}
 \end{eqnarray}
where $N$ is the lapse function, $a$ is the scale factor, and the anisotropy parameters $\beta_i$ satisfy the condition
\begin{equation}
\beta_1+\beta_2+\beta_3=0.
\label{anis}
\end{equation}
Let us suppose that the universe is filled with a spatially homogeneous scalar field, conformally coupled to the scalar curvature. Then
the Klein-Gordon equation (\ref{KG2}) is
\begin{equation}
\ddot{\sigma} + \left(3\frac{\dot{a}}{a}-\frac{\dot{N}}{N}\right)\dot{\sigma} = 0,
\label{KG3}
\end{equation}
where a ``dot'' means the derivative with respect to the time parameter $\tau$.

Let us introduce the shear
\begin{equation}
\theta\equiv \dot\beta_1^2+\dot\beta_2^2+\dot\beta_3^2.
\end{equation}

It is easy~\cite{KPTVV2016BI} to get equation to $\theta$
\begin{equation}\label{Equtheta}
\frac{\dot \theta}{\theta}=2\left[\frac{\dot N}{N}-3\frac{\dot{a}}{a}-\frac{\dot U}{U}\right],
\end{equation}
that can be integrated:
\begin{equation}
\label{theta}
 \theta=\frac{N^2}{U^2a^6}\theta_0,
\end{equation}
where $\theta_0$ is a constant.

One can see that it is convenient to choose the lapse function $N$ as follows~\cite{Star-dipl}:
\begin{equation}
N=a^3.
\label{Star-time}
\end{equation}
In terms of the time parameter, determined by this choice of the lapse function, the Klein-Gordon equation (\ref{KG3}) acquires a particularly simple form:
\begin{equation}
\ddot{\sigma} = 0.
\label{KG4}
\end{equation}
Its solution is a linear function which we shall choose in the following form:
\begin{equation}
\sigma =\sqrt{12U_0}\left(\frac{2\tau}{T}-1\right).
\label{KG5}
\end{equation}
Such a form is convenient because at the moment $\tau = 0$ and $\tau = T$, the scalar $\sigma$ takes such values for which  the function $U$, given by
(\ref{conformal}), is equal to zero.

Using the formulas (\ref{conformal}) and (\ref{Star-time}) we can write down the $00$ component of the Einstein equation (\ref{Einstein1}) as
\begin{equation}
\left(U_0-\frac{\sigma^2}{12}\right)\left(6h^2-\theta\right)-h\dot{\sigma}\sigma=\frac12\dot{\sigma}^2,
\label{Einstein00}
\end{equation}
where $h \equiv \dot{a}/{a}$.
The spatial $ii$ components of the Einstein equation have the following form
\begin{equation*}
\left(U_0-\frac{\sigma^2}{12}\right)\left(\frac{\ddot{a}}{a}-\frac{\dot{a}^2}{a^2}+\ddot{\beta}_i\right)
-\frac{\sigma\dot{\sigma}}{6}\left(\frac{\dot{a}}{a}+\dot{\beta}_i\right)-\frac{1}{12}\dot{\sigma}^2=0.
\end{equation*}
After some algebra we have
\begin{equation}
\left(U_0-\frac{\sigma^2}{12}\right)\ddot{\beta}_1-\frac{\sigma\dot{\sigma}}{6}\dot{\beta}_1 = 0.
\label{beta1}
\end{equation}
The solution of this equation is
\begin{equation}
\dot{\beta_1}=\frac{\beta_{10}}{\left(U_0-\frac{\sigma^2}{12}\right)},
\label{beta11}
\end{equation}
where $\beta_{10}$ is a constant. Analogous solutions can be found for the anisotropy factors $\beta_2$ and $\beta_3$. Finally,
\begin{equation}
\theta = \frac{\theta_0}{\left(U_0-\frac{\sigma^2}{12}\right)^2},\ \theta_0=\beta_{10}^2+\beta_{20}^2+\beta_{30}^2.
\label{theta1}
\end{equation}

Substituting the expressions (\ref{theta1}) and (\ref{KG5}) into Eq.~(\ref{Einstein00}), we obtain
\begin{equation}
h^2+\left(\frac{T-2\tau}{\tau(T-\tau)}\right)h-\frac{1}{\tau(T-\tau)}-\frac{\theta_0T^4}{96U_0^2\tau^2(T-\tau)^2}=0,
\label{h}
\end{equation}

Solving quadratic equation (\ref{h}) with respect to $h$ and choosing the positive sign in front of square root, which corresponds to an expansion, we obtain
\begin{equation}
h=\frac{\tau+\tau_0}{\tau(T-\tau)},
\label{h1}
\end{equation}
with
\begin{equation}
\tau_0=\frac{T}{2}\left(\sqrt{1+\frac{\theta_0T^2}{24U_0^2}}-1\right).
\label{h2}
\end{equation}
Correspondingly, the scale factor behaves as
\begin{equation}
a(\tau)=a_0|\tau^{\frac{\tau_0}{T}}||T-\tau|^{-\frac{\tau_0}{T}-1},
\label{a}
\end{equation}
where $a_0$ is a constant.

We would now like to find the dependence of the cosmic time parameter $t$ on the time parameter $\tau$.
This dependence is given by the integral
\begin{equation}
t = \int N(\tau)d \tau = \int a^3(\tau) d\tau.
\label{cosm-time}
\end{equation}
We shall find this dependence by considering separately three regions of the values of the parameter $\tau$:
\begin{equation}
0 \leq \tau \leq T,
\label{region1}
\end{equation}
\begin{equation}
T \leq \tau < \infty,
\label{region2}
\end{equation}
\begin{equation}
-\infty < \tau \leq 0.
\label{region3}
\end{equation}

In the first region, given by (\ref{region1})
the time parameter is
\begin{eqnarray}
&&t=a_0^3\int d\tau \tau^{3\frac{\tau_0}{T}}(T-\tau)^{-\frac{3\tau_0}{T}-3} \nonumber \\
&&= \frac{a_0^3}{T^2}\left[\frac{1}{k_1}\left(\frac{\tau}{T-\tau}\right)^{k_1}+\frac{1}{k_2}\left(\frac{\tau}{T-\tau}\right)^{k_2}\right],
\label{time1}
\end{eqnarray}
where
\begin{equation*}
\begin{split}
k_1&=\frac32\sqrt{1+\frac{\theta_0T^2}{24U_0^2}}-\frac12=3\frac{\tau_0}{T}+1,\\
k_2&=\frac32\sqrt{1+\frac{\theta_0T^2}{24U_0^2}}+\frac12=3\frac{\tau_0}{T}+2.
\end{split}
\end{equation*}
The integration constant in (\ref{time1}) is chosen in such a way that when $\tau \rightarrow 0$, the cosmic time $t$ also tends to zero and this moment corresponds to
the Big Bang singularity as one can see from the formulas (\ref{h1}) and (\ref{a}). When $\tau \rightarrow T$, the cosmic time tends to $+\infty$ and the universe undergoes an infinite expansion.   We can now study the dependence of the scale factor $a$ on cosmic time $t$ in these two asymptotic cases.

First, when $\tau \rightarrow 0$:
\begin{eqnarray}
&&t \sim \tau^{k_1},\quad \Rightarrow \quad \tau \sim   t^{\frac{1}{k_1}},\nonumber \\
&&a \sim \tau^{\frac{\tau_0}{T}} \sim t^{\frac{\tau_0}{k_1T}} = t^{\frac{\sqrt{1+\frac{\theta_0T^2}{24U_0^2}}-1}{3\sqrt{1+\frac{\theta_0T^2}{24U_0^2}}-1}}=t^{\frac{\tau_0}{3\tau_0+T}}.
\label{time1-as}
\end{eqnarray}
One can easily see that this formula corresponds to the formulas (60) and (61), obtained in  \cite{we-Bianchi} on analysing  the Einstein equations in the vicinity of the singularity
in terms of the cosmic time $t$.

When $\tau \rightarrow T$ and $t \rightarrow \infty$, we have
\begin{eqnarray}
&& t \sim \frac{1}{(T-\tau)^{k_2}}\quad \Rightarrow \quad T-\tau \sim t^{-\frac{1}{k_2}},\nonumber \\
&&a \sim t^{\frac{\sqrt{1+\frac{\theta_0T^2}{24U_0^2}}+1}{3\sqrt{1+\frac{\theta_0T^2}{24U_0^2}}+1}}=t^{\frac{\tau_0+T}{3\tau_0+2T}}.
\label{time1-as1}
\end{eqnarray}
This asymptotic behavior also coincides with that described in \cite{we-Bianchi}.
One can easily also check  that the behavior of the anisotropy factor is Kasner-like and coincides with that found
in \cite{we-Bianchi}. Indeed,
\begin{equation}
\dot{\beta}_i = \frac{\beta_{i0}}{U_0-\frac{\sigma^2}{12}}=\frac{\beta_{i0} T^2}{4U_0\tau(T-\tau)}.
\label{Kasner}
\end{equation}
In both the asymptotic regimes, $\tau \rightarrow 0$ and $\tau \rightarrow T$ the integration of Eq.~(\ref{Kasner}) gives
$\beta_i \sim \ln \tau$ or $\beta_i \sim \ln (T-\tau)$, which, in turn implies the power-law dependence of the factors $e^{\beta_i}$ on the cosmic time $t$.

 As a matter of fact, it would be more correct to speak about the generalized Kasner-like behavior, and, correspondingly, about the generalized Kasner indices.
We can also write down  the corresponding relations for these indices in terms of  the quantities, introduced in this paper. Thus, introducing these generalized Kasner indices 
$p_i$ through the relation 
\begin{equation}
ae^{\beta_i} \sim t^{p_i},
\label{gen-Kas}
\end{equation}
we see that for $t \rightarrow 0$ we have the following pair of relations:
\begin{equation}
p_1+p_2+p_3 = \frac{3\tau_0}{3\tau_0+T},
\label{gen-Kas1}
\end{equation}
\begin{equation}
p_1^2+p_2^2+p_3^2 = \frac{3\tau_0(3\tau_0+2T)}{(3\tau_0+T)^2}.
\label{gen-Kas2}
\end{equation}
Instead  for $t \rightarrow \infty$ the generalized Kasner indices satisfy the relations
\begin{equation}
p_1+p_2+p_3 = \frac{3(\tau_0+T)}{3\tau_0+2T},
\label{gen-Kas3}
\end{equation}
\begin{equation}
p_1^2+p_2^2+p_3^2 = \frac{3(\tau_0+T)(3\tau_0+T)}{(3\tau_0+2T)^2}.
\label{gen-Kas4}
\end{equation}
In both cases, 
$$
\sum_{i=1}^3 p_i^2=2\sum_{i=1}^3 p_i -\left(\sum_{i=1}^3 p_i\right)^2.
$$
Let us note that  while the standard Kasner indices \cite{Kasner} form a one-parameter family, which can be parametrized, for example, by the Lifshitz-Khalatnikov parameter \cite{Khal-Lif}, in the formulas (\ref{gen-Kas1})-(\ref{gen-Kas2}) and in the formulas (\ref{gen-Kas3})-(\ref{gen-Kas4}) one encounters two-parameter families.
One could remember that a two-parameter family of the Kasner indices was already studied for the case of a Bianchi-I universe filled with a minimally coupled massless 
scalar field \cite{Bel-Khal} (see also the section III of the paper \cite{we-Bianchi}). However, in that case the sum of the three Kasner indices was still equal to 1. Moreover, the relations for the Kasner indices were valid during all the cosmological evolution. 

We wish now  to consider the parametric time interval (\ref{region2}). Here
\begin{eqnarray}
&&t = a_0^3\int d\tau \tau^{3\frac{\tau_0}{T}}(\tau-T)^{-\frac{3\tau_0}{T}-3} \nonumber \\
&& =\frac{a_0^3}{T^2}\left[\frac{1}{k_1}\left(\frac{\tau}{\tau-T}\right)^{k_1}-\frac{1}{k_2}\left(\frac{\tau}{\tau-T}\right)^{k_2}\right].
\label{time2}
\end{eqnarray}
When $\tau$ is close to $T$ the cosmic time (\ref{time2}) tends to $-\infty$ and the cosmic scale factor $a$ grows indefinitely. Then for $\tau \rightarrow +\infty$, the cosmic time tends to a finite constant value $t_0$, where
\begin{equation}
t_0 = \frac{32a_0^3U_0^2}{T^2\left(3\theta_0T^2+64U_0^2\right)}.
\label{time0}
\end{equation}
Thus, the universe undergoes an infinite contraction which begins at $t = -\infty$ and ends at $t=t_0$. Let us see how the scale factor behaves asymptotically.
At $\tau \rightarrow T_+$ we have
\begin{eqnarray}
&&t \sim -\frac{1}{(\tau-T)^{k_2}} \quad \Rightarrow \quad (\tau-T) \sim \left(-\frac{1}{t}\right)^{\frac{1}{k_2}},\nonumber \\
&&a \sim (-t)^{\frac{\sqrt{1+\frac{\theta_0T^2}{24U_0^2}}+1}{3\sqrt{1+\frac{\theta_0T^2}{24U_0^2}}+1}}=(-t)^{\frac{\tau_0+T}{3\tau_0+2T}}.
\label{time-2as}
\end{eqnarray}
On the other hand, when $\tau \rightarrow +\infty$
\begin{eqnarray}
&&(t-t_0) \sim {}-\frac{1}{\tau^2}\quad \Rightarrow \quad \tau \sim \frac{1}{\sqrt{t_0-t}},\nonumber \\
&&a \sim \frac{1}{\tau} \sim \sqrt{t_0-t}.
\label{time2-as1}
\end{eqnarray}
As far as the anisotropy factors are concerned, when $\tau$ is close to $T$ and the universe is infinitely large, we again have a Kasner-like behavior. The situation changes when we are close to the Big Crunch singularity, i.e. when $\tau \rightarrow +\infty$. Indeed, in this case the integration
of the formula (\ref{Kasner}) gives $\beta_i \sim 1/\tau \rightarrow 0$ and $e^{\beta_i}$ does not reveal a power-law behavior. Instead, one has some kind of the isotropization, but in contrast to the isotropization in the Heckmann-Schucking solution for the Bianchi-I universe filled with dust \cite{Heck-Schuck,Khal-Kam}, here the isotropization occurs during the contraction of universe.

Finally, in the third region  (\ref{region3})
\begin{eqnarray}
&&t = a_0^2\int d\tau (-\tau)^{3\frac{\tau_0}{T}}(T-\tau)^{-\frac{3\tau_0}{T}-3} \nonumber \\
&& = \frac{a_0^3}{T^2} \left[\frac{1}{k_2}\left(\frac{-\tau}{T-\tau}\right)^{k_2}-\frac{1}{k_1}\left(\frac{-\tau}{T-\tau}\right)^{k_1}\right].
\label{time3}
\end{eqnarray}
Now, when $\tau \rightarrow 0_-$, also the cosmic time $t \rightarrow 0_-$ and
\begin{eqnarray}
&&t \sim -(-\tau)^{k_1}\quad \Rightarrow \quad \tau \sim -(-t)^{\frac{1}{k_1}},\nonumber \\
&& a \sim  (-t)^{\frac{\sqrt{1+\frac{\theta_0T^2}{24U_0^2}}-1}{3\sqrt{1+\frac{\theta_0T^2}{24U_0^2}}-1}}=(-t)^{\frac{\tau_0}{3\tau_0+T}}.
\label{time3-as}
\end{eqnarray}
When $\tau \rightarrow -\infty$, the cosmic time $t$ tends to a finite value $t_1={}-t_0$.
In the vicinity of this point
\begin{equation}
a \sim \sqrt{t+t_0}.
\label{time3-as1}
\end{equation}
Let us note also that as follows from Eq.~(\ref{h1}), at the moment when the parametric time is equal to $\tau = -\tau_0$, the Hubble parameter is equal to zero and the
time derivative of the scale factor is equal to zero. This moment correspond to the point of the maximal expansion of the universe.
The corresponding cosmic time  value is
\begin{equation}
t_1=\frac{a_0^3}{T^2}\left[\frac{1}{k_2}\left(\frac{\tau_0}{T+\tau_0}\right)^{k_2}-\frac{1}{k_1}\left(\frac{\tau_0}{T+\tau_0}\right)^{k_1}\right].
\label{time100}
\end{equation}

Let us note, that  here at the Big Bang $\tau \rightarrow -\infty$, the anisotropy is absent, while the universe behaves in a  generalized Kasner-like way when the universe arrives at the Big Crunch singularity at $\tau \rightarrow 0_-$.

 Finally, let us note that from the traditional point of view only the interval with the parametric time $\tau$ confined in the interval 
$0 \leq \tau \leq T$, where  gravity has a positive sign, has a physical sense \cite{Star-dipl}. Nevertheless, in the next section we shall try to describe 
the evolution of the universe,  including all the three intervals of the parametric time, by making some kind of matching, then implementing   some prescriptions for the singularity crossings.

\section{Complete evolution, singularity crossings and the relation between the Jordan frame and the Einstein frame}

We are now in a position to describe the complete evolution of the Bianchi-I universe filled with the massless scalar field conformally coupled to gravity, including the singularity crossings. The latter will be treated in the spirit of the approach developed in our preceding publications \cite{KPTVV-sing,we-Bianchi}.

As  follows from the preceding section, the first stage of the evolution corresponds to the second region of the parametric time running (\ref{time2}). It  begins in the infinitely remote past $t \rightarrow -\infty$ when the volume of the universe is also infinite and it begins its contraction. The universe is anisotropic and its behavior is characterized by some set of Kasner indices. The process of contraction occupies an infinite cosmic time and at the moment $t = t_0$ it encounters the Big Crunch singularity. As  discussed above (see Eq.~(\ref{time2-as1}) ) the scale factor in this limit behaves as $a \sim \sqrt{t_0-t}$ and the anisotropy factors disappear. Thus, we encounter in the vicinity  of this Big Crunch singularity the situation, which was already studied in \cite{KPTVV-sing}, where the evolution of the flat Friedmann universe filled with a massless scalar field was considered. In paper~\cite{KPTVV-sing} the transformations between the Jordan and the Einstein frames were studied in detail.
As is well-known  the action, including the scalar field, non-minimally coupled to the gravity can be transformed into the action with a minimally coupled canonically normalized scalar field by means of the conformal transformation of the metric, combined with a certain transformation of the scalar field. Such a combination of transformations is usually called a transition between the Jordan frame and the Einstein frame. The form of these transformations is  particularly convenient, when we consider the case of the conformal coupling, because  in this case these transformations are invertible. Indeed, for  values of  the scalar field $\sigma$  such that
$-\sqrt{12U_0} \leq \sigma \leq \sqrt{12U_0}$, or, in terms of the preceding section, if $ 0 \leq \tau \leq T$, then the transformations have the following form:
\begin{equation}
\tilde{g}_{\mu\nu} = g_{\mu\nu}\frac{U_0-\frac{\sigma^2}{12}}{U_1},
\label{JE}
\end{equation}
\begin{equation}
\phi = \sqrt{3U_1}\ln \left[\frac{\sqrt{12U_0}+\sigma}{\sqrt{12U_0}-\sigma}\right],
\label{JE1}
\end{equation}
\begin{equation}
\sigma = \sqrt{12U_0}\tanh \left[\frac{\phi}{\sqrt{12U_1}}\right].
\label{JE2}
\end{equation}
Here, $\tilde{g}_{\mu\nu}$ is a new metric, $\phi$ is a new scalar field and $U_1$ is a positive constant.
In terms of these new variables the coupling between the scalar field and gravity becomes minimal.

However, if we consider the values of the scalar field $\sigma$ such that $|\sigma| > \sqrt{12U_0}$, then
the situation changes. The constant $U_1$ changes sign (this effect could be called ``antigravity'') and
the new scalar field $\phi$ should be substituted by the phantom scalar field $\chi$, whose kinetic term has a negative
sign. The relation between the fields $\sigma$ and $\chi$ is given by the formula
\begin{equation}
\sigma = \sqrt{12U_0}\coth \left[\frac{\chi}{\sqrt{12|U_1|}}\right].
\label{JE3}
\end{equation}

In paper \cite{KPTVV-sing} it was shown that for the Friedmann universe, when one encounters the cosmological singularity in the Einstein frame,
the evolution is regular in the Jordan frame and vice versa. This fact was used to describe the Big Bang - Big Crunch singularity crossing.
We can now come back to our Bianchi-I universe. As  was explained above, when the infinite contraction  ends in the encounter with the Big Crunch singularity,
the universe behaves like an isotropic one and its contraction is described by the formula $a \sim \sqrt{t_0-t}$. At the same time, when our parametric time changes  in the interval (\ref{time2}), the corresponding evolution begins with  the Big Bang singularity and the expansion is described by the formula $a \sim \sqrt{t-(-t_0)}$. At this point, it is convenient to change the integration constant in the formula (\ref{time2}) adding to the right-hand side of this equation the constant $-2t_0$. The Big Crunch and Big Bang are now labeled by the same cosmic time moment $t = -t_0$. Moreover, on applying the method, described in paper \cite{KPTVV-sing}, one can see that these singularities
disappear if we make the transition to the Einstein frame. Thus, on using the correspondence formulas between the frames, one can describe the Big Crunch - Big Bang singularity crossing in the model with the conformally coupled scalar field. Namely, the parametric time moment $\tau = +\infty$ is matched with the parametric time moment
$\tau = -\infty$ and both correspond to the cosmic time moment $t = -t_0$. What happens to the scalar field  $\sigma$ in the moment of the singularity crossing?
If we apply the formula (\ref{KG5}), we shall see that its value  changes from $\sigma = +\infty$ to $\sigma = -\infty$. This jump appears to be quite natural from the point of view of the regular evolution in the Einstein frame. Indeed,   in the Einstein frame at the moment of the  singularity crossing the phantom field $\chi$ is equal to zero.
On passing through zero, the phantom field changes  sign and it follows from the formula (\ref{JE3}) that the scalar field $\sigma$ undergoes and infinite jump.

After the crossing of this isotropic Big Crunch - Big Bang singularity, the evolution of the universe corresponds to the changing of the parametric time $\tau$ from $\tau =-\infty$ to $\tau = 0$. During this evolution the cosmic time initial value, corresponding to the Big Bang is equal to $t = -t_0$, the universe  expands until the moment
$\tau = -\tau_0,\  t = t_1$, corresponding to the maximal scale factor, then it begins a contraction, culminating in the second encounter with the Big Crunch singularity at
$\tau = 0,\  t = 0$. However, this Big Crunch singularity is different from that encountered at $t=-t_0$. As was already mentioned above the universe at $\tau \rightarrow 0$ has a Kasner-like behavior and is essentially anisotropic. This singularity occurs simultaneously in both the Jordan and the Einstein frames \cite{we-Bianchi} and the regularity of the evolution in one frame cannot be used for the description of the singularity crossing in another frame. In paper \cite{we-Bianchi} a very simple method for the description of the singularity crossing for some cosmological models was suggested. It was based on the introduction of some new variables, defined at certain regions of the phase space with the subsequent analytical extension of the applicability of the laws of the evolution of these variables outside of their initial domains.  This method gives the same prescription for the singularity crossing of the Friedmann universe as  other methods  and is easily applicable also for the Bianchi-I universe, filled with minimally coupled scalar field. Here, in spite of the fact that in both the frames we have a singularity, on knowing what happens in the Einstein frame, we can understand how the universe crosses the Big Crunch -- Big Bang singularity in the Jordan frame. One can see that  the  behavior of the scale factor $a$ at $t \rightarrow 0_+$, given by the formula (\ref{time1-as}), is quite similar to the behavior of the scale factor of the universe at $t \rightarrow 0_-$, given by the formula (\ref{time3-as}), and these formulas can be unified given the description of the
Big Crunch - Big Bang singularity as
\begin{eqnarray}
&& a \sim  |t|^{\frac{\sqrt{1+\frac{\theta_0T^2}{24U_0^2}}-1}{3\sqrt{1+\frac{\theta_0T^2}{24U_0^2}}-1}}=|t|^{\frac{\tau_0}{3\tau_0+T}}.
\label{time3-as0}
\end{eqnarray}
Let us note that the behavior of the scalar field $\sigma$ at the crossing of this singularity is quite regular, it is linearly growing with the time parameter $\tau$ passing through the value $\sigma = -\sqrt{12U_0}$.

\begin{figure}[t]
\centerline{\epsfxsize 8cm
\epsfbox{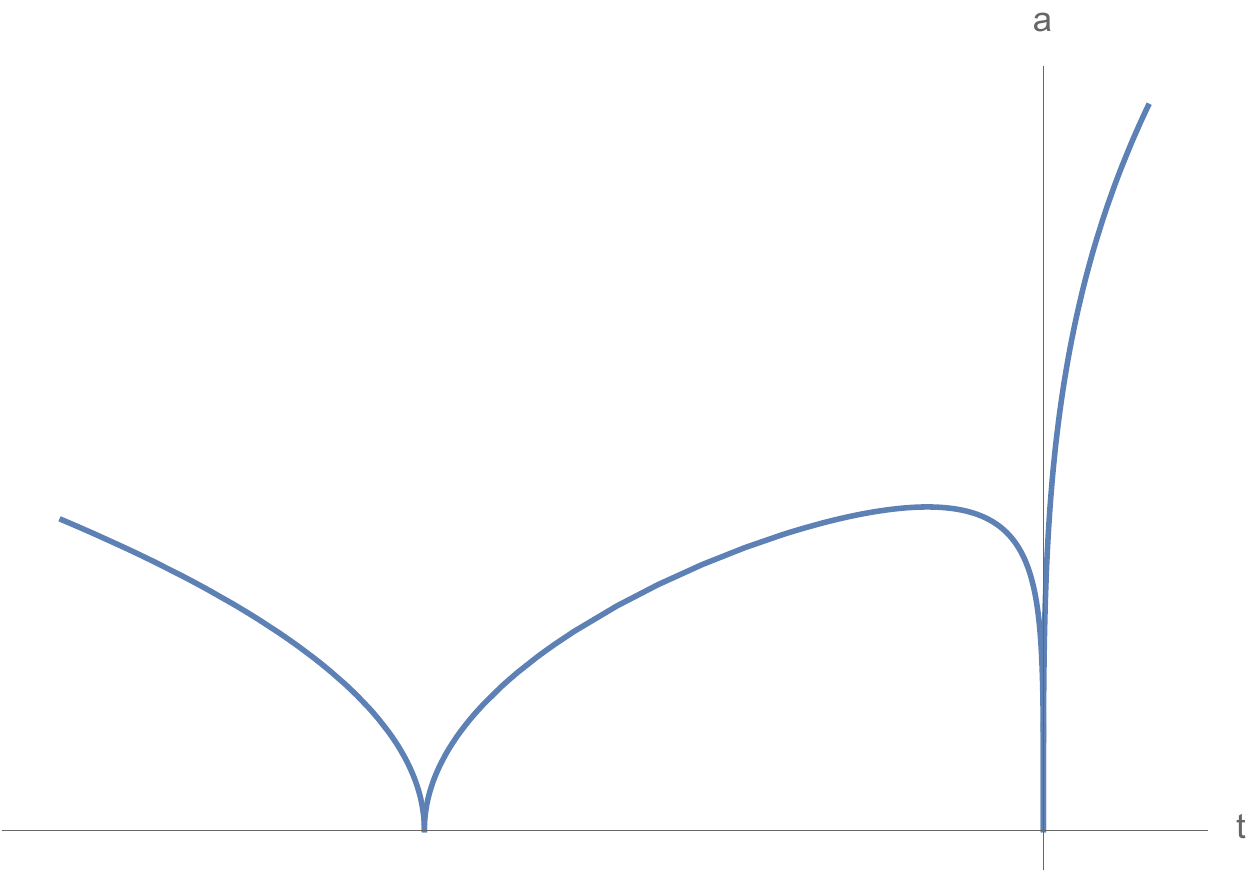}}
\caption{Cosmological evolution of the Bianchi-I universe filled with a conformally coupled scalar field.}
\label{conf}
\end{figure}

\begin{figure}[t]
\centerline{\epsfxsize 8cm
\epsfbox{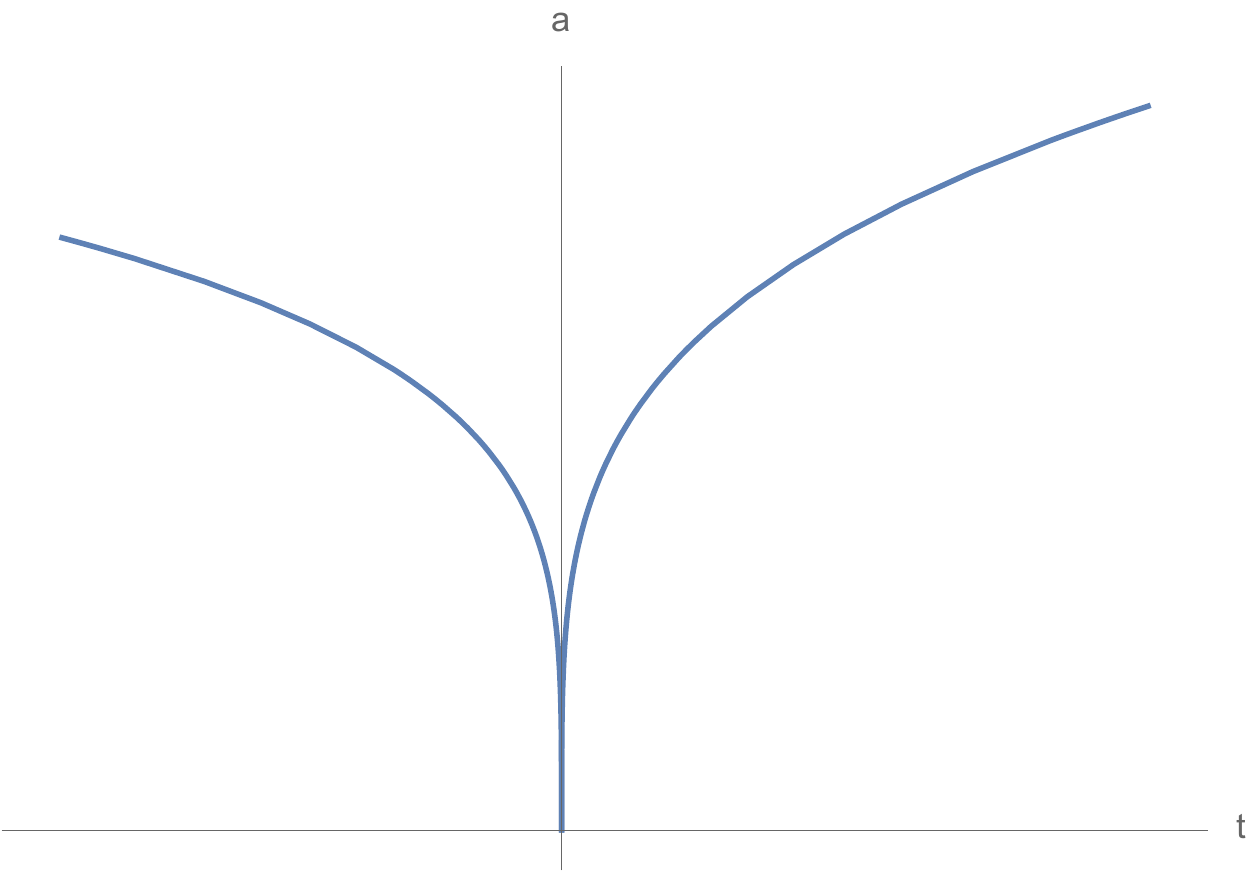}}
\caption{Cosmological evolution of the Bianchi-I universe filled with a conformally coupled scalar field -- crossing of the singularity at $t=0$.}
\label{conf1}
\end{figure}

Finally, after this second Big Bang, the universe begins an infinite expansion, when the cosmic time  runs from $t=0$ to $t=\infty$. On summing up, we can say that the universe begins its evolution being infinitely large and contracting. Then it crosses the isotropic Big Crunch - Big Bang singularity and begins expanding. This expansion stops at some moment of time and the universe undergoes a period of contraction culminating in the encounter with the anisotropic Big Bang - Big Crunch singularity. Having crossed this singularity the universe begins an infinite expansion. The evolution of the Bianchi-I universe filled with a conformally coupled scalar field is presented in Figures \ref{conf} and \ref{conf1}.

\begin{figure}[t]
\centerline{\epsfxsize 8cm
\epsfbox{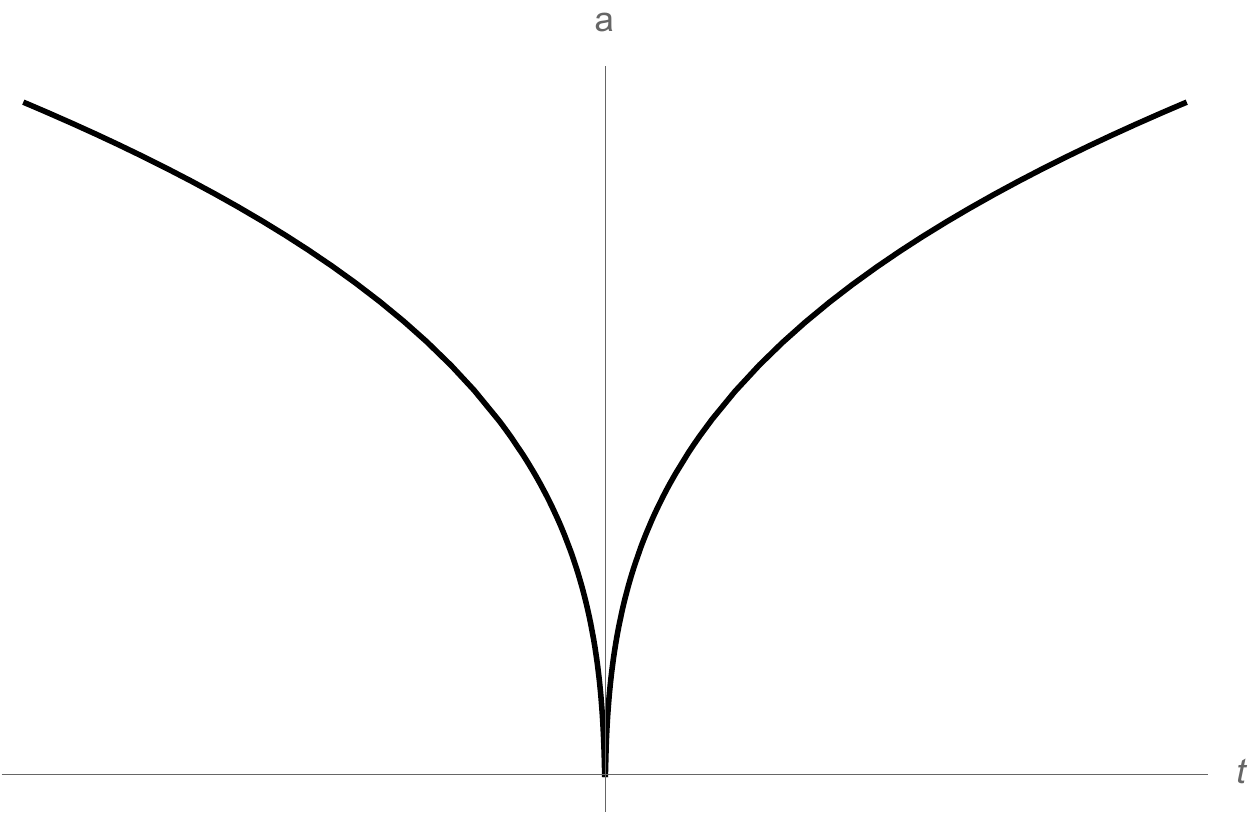}}
\caption{Cosmological evolution of the Bianchi-I universe filled with a minimally coupled scalar field}.
\label{min}
\end{figure}

It is interesting  to compare the evolution of the universe in the Bianchi-I universe, filled with a conformally coupled scalar field  with that, filled with a minimally coupled scalar field. The latter  is much simpler. It is known  that the corresponding equations of motion can be solved in terms of the cosmic time parameter $t$ and
there are two solutions: an infinite expansion which begins from the Big Bang singularity and an infinite contraction culminating in the encounter with the Big Crunch singularity. 
The crossing of the Big Crunch -- singularity was described in paper \cite{we-Bianchi} and the complete evolution is given by the formula
\begin{equation}
a(t) = a_0|t|^{\frac13},
\label{min-complete}
\end{equation}
where $t$ runs from $-\infty$ to $+\infty$. In contrast with the case of the conformally coupled scalar field,  we here have only one anisotropic Big Crunch -- Big Bang singularity (see Fig.~\ref{min}).

 In all the above, we have concentrated on the description of the matching between the behaviors of the scale factor $a$ before and after the singularity. Indeed, from our point of view, this question is more complicated when the behavior of the anisotropy functions $\beta_i$ is simpler. As we know the  time derivative of these functions 
is inversely proportional to the cubes of the scale factors. Thus, on fixing the proportionality constant, we can fix the behavior of the anisotropy functions, provided the rules 
for the continuation of the scale factor are defined. One should remember that, integrating the equations for the anisotropy functions, one also obtains  an additive constant. We can fix the value of this constant as well. 

Let us remember that there is an intensive discussion in the literature concerning the equivalence between the Jordan and the Einstein frames in classical and quantum cosmology
(see e.g. \cite{Christian,Christian1} and references therein). In our preceding papers \cite{KPTVV2013,KPTVV2015,KPTVV-sing,we-Bianchi} we have expressed the opinion that while from the mathematical point of view these two frames are equivalent, the physical pictures can be quite different. We believe that the presented picture of the cosmological evolution in Bianchi-I models gives a nice illustration of our point of view.

Concluding we wish to say that the very topic of the crossing of the Big Bang - Big Crunch singularity is still rather controversial. However, we believe that it is worth  further studies. Perhaps, the status of the field reparametrizations, allowing one  to go beyond singularities is similar to the  changing of coordinates in the description of black holes \cite{Kruskal}. The latter does not remove the horizons, but allows us to see what happens behind the horizons. We think that something similar can also take place  for cosmological singularities.

\section{Induced gravity and Bianchi - I universe}

In this section we present the general exact solution for the Bianchi - I universe in the induced gravity model. 
The induced gravity model corresponds to the following choice of the coupling function between the scalar field and the scalar curvature:
\begin{equation}
U(\sigma) = \frac12\gamma\sigma^2,
\label{induced}
\end{equation}
where the constant $\gamma$ is positive. The induced gravity was first suggested in paper \cite{Sakharov} and then has found many applications in cosmology
\cite{ind-cosm,ind-cosm1,ind-cosm2,ind-cosm3}. The general relations between the induced gravity models and their counterparts with minimally coupled scalar fields were considered in papers \cite{KPTVV2013,KPTVV2015}.  Substituting the function (\ref{induced}) and the metric (\ref{Bianchi}) into Eqs. (\ref{Einstein}) and (\ref{KG}), we obtain the following 
system of equations:
\begin{equation}
3\gamma\sigma^2\frac{\dot{a}^2}{a^2}-\frac12\gamma\sigma^2\theta+6\gamma\dot{\sigma}\sigma\frac{\dot{a}}{a}-\frac12\dot{\sigma}^2=0,
\label{induced1}
\end{equation}
\begin{eqnarray}
&&-\gamma\sigma^2\frac{\ddot{a}}{a}-\frac12\gamma\sigma^2\frac{\dot{a}^2}{a^2}+\gamma\sigma^2\frac{\dot{N}\dot{a}}{Na}-\frac14\gamma\sigma^2\theta\nonumber \\
&&-\gamma\ddot{\sigma}\sigma+\gamma\dot{\sigma}\sigma\frac{\dot{N}}{N}-2\gamma\dot{\sigma}\sigma\frac{\dot{a}}{a}-\gamma\dot{\sigma}^2-\frac14\dot{\sigma}^2=0,
\label{induced2}
\end{eqnarray}
\begin{eqnarray}
&&\ddot{\sigma}-\frac{\dot{N}}{N}\dot{\sigma}+3\frac{\dot{a}}{a}\dot{\sigma}-6\gamma\sigma\frac{\ddot{a}}{a}+6\gamma\sigma\frac{\dot{N}\dot{a}}{Na}\nonumber \\
&&-6\gamma\sigma\frac{\dot{a}^2}{a^2}-\gamma\sigma\theta=0.
\label{induced3}
\end{eqnarray}
Writing down these equations, we have also taken into account the fact that 
\begin{equation}
\beta_i=\frac{\beta_{i0}N}{a^3U}, 
\label{induced4}
\end{equation}
for any choice of the lapse function $N$ and the coupling function $U$.

Let us now multiply Eq. (\ref{induced1}) by $\frac{1}{\sigma}$ and Eq. (\ref{induced2}) by $-\frac{6}{\sigma}$ and add both of them to Eq. (\ref{induced3}).
As a result we obtain 
\begin{equation}
\ddot{\sigma}-\frac{\dot{N}}{N}\dot{\sigma}+3\frac{\dot{a}}{a}\dot{\sigma}+\frac{\dot{\sigma}^2}{\sigma}=0.
\label{induced5}
\end{equation}

On now choosing  the same time parameter $\tau$, as that used in the Section III, which corresponded to the lapse function $N = a^3$, be obtain a very simple equation
\begin{equation}
\ddot{\sigma}+\frac{\dot{\sigma}^2}{\sigma}=0.
\label{induced6}
\end{equation}
The general solution of this equation is
\begin{equation}
\sigma = \sigma_0\sqrt{\tau},
\label{induced7}
\end{equation}
where one of the two arbitrary constants, parametrizing  this solution, is included into the choice of the initial moment of $\tau$. Substituting the solution (\ref{induced7}) 
into Eq. (\ref{induced1}) and taking into account also the expression (\ref{theta}), we see that $\frac{\dot{a}}{a}$ is proportional to $\frac{1}{\tau}$. That means that 
the scale factor $a$ is a power-law function of $\tau$. Moreover, it immediately follows  from Eq. (\ref{cosm-time}) that the cosmic time is also a power-law function of the parametric time $\tau$. As a result, both the scale factor $a$ and the scalar field $\sigma$ can be found as a power-law functions of the cosmic time $t$.

Thus, let us look for the solution of Eqs. (\ref{induced1})--(\ref{induced3}) in the following form:
\begin{eqnarray}
\sigma(t) = \sigma_0t^{s},\\
a(t) = a_0t^{r}.
\label{induced8}
\end{eqnarray}
Substituting the formulas (\ref{induced8}) into Eq. (\ref{induced1}) we obtain the following relation between the exponents $r$ and $s$:
\begin{equation}
3r+2s = 1,
\label{induced9}
\end{equation}
or
\begin{equation}
s=\frac12-\frac32r.
\label{induced10}
\end{equation}

Another relation following from Eq. (\ref{induced1}) and from the relation (\ref{induced10}) is 
\begin{equation}
-r^2(16\gamma+3)+2r(4\gamma+1)-\frac13=\frac{16\theta_0}{3a_0^6\sigma_0^4\gamma}.
\label{induced11}
\end{equation}
It immediately follows  from the last equation that at a fixed value of the coefficient $\gamma$ the choice of the exponent $r$ is limited by the inequality 
\begin{eqnarray}
&&\frac{4\gamma+1}{16\gamma+3}-\frac{1}{16\gamma+3} \sqrt{\frac{8\gamma(6\gamma+1)}{3}}\leq r \nonumber \\
&&\leq \frac{4\gamma+1}{16\gamma+3}+\frac{1}{16\gamma+3} \sqrt{\frac{8\gamma(6\gamma+1)}{3}}.  
\label{induced12}
\end{eqnarray}
We see that the acceptable values of the exponent $r$ are always positive and that they are less than $\frac12$.  Thus, when we consider positive values of the cosmic time parameter $t$, we 
always have an expansion. If we wish to consider a contracting universe, we should substitute in the formulas (\ref{induced8}) $t$ by $(-t)$, where  $-\infty < t \leq 0$. Note that in the values of $r$, satisfying the inequalities (\ref{induced12}) can be less or greater than $\frac13$. The parameter $s$ can be positive or negative respectively. This means that when the universe  approaches the singularity at $t \rightarrow 0$, the scalar field $\sigma$ can grow indefinitely if $s< 0$  and can tend to zero if $s >0$. It is interesting to note   that  a regime, when the scalar field $\sigma$ as well as the effective Newton constant are constant,      
is also possible. That happens when $s=0$, $r= \frac13$. It is easy to see from Eq. (\ref{induced11}) that this happens when 
\begin{equation}
6\theta_0=a_0^6\sigma_0^4\gamma^2.
\label{induced13}
\end{equation}
Such a regime is absent in the case of the isotropic flat Friedmann model, i.e. when $\theta_0 = 0$. 

It is interesting now to see which regime corresponds to that, described by the formulas (\ref{induced8}), (\ref{induced9}) when we make the transformation from the Jordan frame to the Einstein frame. As  described in our preceding papers \cite{KPTVV2013,KPTVV2015}, such a transformation gives the following relation between the cosmological radii in these frames 
\begin{equation}
\tilde{a} = a\sqrt{\frac{U}{U_1}}.
\label{J-E}
\end{equation}
For the case of the induced gravity this means that 
\begin{equation}
\tilde{a} \sim a\times \sigma \sim t^{r+s}.
\label{J-E1}
\end{equation}
However, we want to express the scalar factor $\tilde{a}$ in the Einstein frame as a function of the cosmic time $\tilde{t}$ in the Einstein frame.
Thus, we have
\begin{equation}
\tilde{t} = \int dt \tilde{N} \sim \int dt \sigma(t) \sim \int dt t^{s} \sim t^{s+1}.
\label{J-E2}
\end{equation}
Substituting the expression (\ref{J-E2}) into Eq. (\ref{J-E1}) and taking into account the relation (\ref{induced9}), we obtain
\begin{equation}
\tilde{a} \sim \tilde{t}^{\frac{r+s}{s+1}} \sim \tilde{t}^{\frac13},
\label{J-E3}
\end{equation}
i.e. in the Einstein frame the standard regime (\ref{J-E3}) corresponds to the regime (\ref{induced8})  independently of the value of the 
exponent $r$. Thus, the the singularities in both the frames arise simultaneously and there is no opportunity of describing the singularity crossing in one frame,
using a regular evolution in another frame. We shall not  dwell here   on another suggestion for the description of the singularity crossing, which was presented in
paper \cite{we-Bianchi} and in the preceding section of the present paper. 

We would like now to describe the generalized Kasner regime of the exact solution for the Bianchi-I universe.     
By using the formulas (\ref{induced4}), (\ref{induced7}) and (\ref{induced8}) we see that 
\begin{equation}
\dot{\beta}_i = \frac{2\beta_{i0}}{\gamma\sigma_0^2a_0^3}\frac{1}{t}. 
\label{Kasner-ind}
\end{equation}
Integrating this equation, we can obtain the expressions for the generalized Kasner indices:
\begin{equation}
p_i = r +  \frac{2\beta_{i0}}{\gamma\sigma_0^2a_0^3}.
\label{Kasner-ind1}
\end{equation}
Now, using the relation (\ref{induced11}), we can obtain the relations for the generalized Kasner indices:
\begin{equation}
p_1+p_2+p_3 = 3r,
\label{Kasner-ind2}
\end{equation}
\begin{equation}
p_1^2+p_2^2+p_3^2=1-\frac{(1+4\gamma)(3r-1)^2}{4\gamma}.
\label{Kasner-ind3}
\end{equation}
Let us note that while the sum of the generalized Kasner indices can be greater or less than $1$, the sum of the squares of 
these indices is always less or equal to $1$.

\section{Concluding remarks}
We presented in this paper   exact solutions for the Bianchi-I universe filled with a minimally coupled massless scalar field, a conformally coupled 
massless scalar field and for the induced gravity Bianchi-I model. Furthermore, we discussed some questions, connected with the possible description 
of the Big Bang - Big Crunch cosmological singularities crossing in anisotropic spacetimes. As is well-known the question of the singularity crossing leads to 
some discussions and the situation is far  from  clear. Nonetheless, we believe that the presentation of some non yet conventional ideas can be useful  
and fruitful. The main idea, used in this paper and in the preceding papers, consists of the attempt to find such a parametrization of fields, participating in the cosmological 
evolution, which provides a sufficient number of finite characteristics, which permit one  to describe the matching of the cosmological evolution before and after the singularity crossing. To find such a parametrization, we utilize the presence of a scalar field in the models under consideration. These two features: the role of the scalar field and 
the definition of the finite characteristics of the evolution makes our approach in a way similar to other approaches, sketched in the Introduction of the present paper. 
One should recognize that neither us nor anyone else (at least,  to our knowledge) has a universal recipe and the problem should be analized  in a case by case manner.    

Of course, since curvature invariants strongly diverge in the singularity,
quantum-gravitational effects become important at least at the Planck
curvature (and may be even at a smaller one). Thus, our hypothesis can be
reformulated as the assumption that the finite characteristics we used for
matching of solutions on both sides of the singularity remain mostly
unaffected by these effects, as well as by other possible modifications of
gravity in the region of super-high curvature.

Before concluding we would like to cite one more recent approach to the treatment of the anisotropic cosmological singularities \cite{Krasnov}. In the framework of this approach a family of modified gravity theories is considered. The main role in these theories is played by the fundamental connection. It is shown that in the Kasner 
universe, this connection is always finite even when the metric is singular.   While the authors of \cite{Krasnov} and other papers, developing this idea work with the modified gravity in contrast to our more conservative approach, we find their idea of the finite quantities, existing in all the spacetime attractive and interesting. 

\section*{Acknowledgments}
A.~S. was supported by the RSF grant 16-12-10401.
Research of E.P. is supported in part by grant MK-7835.2016.2  of the President of Russian Federation.
Research of S.V. is supported in part by grant NSh-7989.2016.2 of the President of Russian Federation.

\end{document}